# Microstructure-dependent oxidation-assisted dealloying of $Cu_{0.7}Al_{0.3}$ thin films

Jiangbin Su[1,2,*], Meiping Jiang[1], Honghong Wang[1], Yang Liu[1]

*1. Experiment Center of Electronic Science and Technology & Department of Electronic Science and Engineering, School of Mathematics and Physics, Changzhou University, Changzhou 213164, PR China*

*2. China-Australia Joint Laboratory for Functional Nanomaterials & Physics Department, Xiamen University, Xiamen 361005, PR China*

[*] Corresponding author. Tel.: +86 051986330300. E-mail address: jbsu@cczu.edu.cn (J.B. Su)

**Abstract:** In this paper, the oxidation-assisted dealloying (OAD) of $Cu_{0.7}Al_{0.3}$ films with different microstructures which were obtained by high vacuum annealing at different temperatures were studied using powder X-ray diffraction, field-emission scanning electron microscopy and energy dispersive X-ray analysis. It was observed that different microstructures such as eutectic mixture or solid solution, the grain size of Cu or Al component in eutectic-mixture $Cu_{0.7}Al_{0.3}$ films affected the corrosion morphology greatly. It thus provided a practical route to fabricate flexible CuO porous nanostructure-films (PNFs) with controllable pore size, porosity, block size and shape. Further, the underlying OAD mechanisms for the structure-resultant different corrosion morphologies of $Cu_{0.7}Al_{0.3}$ films were also explored. In these senses, the study is suggestive and crucial to both the mechanism understanding of OAD process and the technical controlling of PNF fabrication.

# 1. Introduction

Porous nanostructure-films (PNFs) [1] are a special kind of porous thin films composed of many nanosized blocks such as nanoparticles, nanoligaments, nanoplates, nanovoids or nanochannels whose dimensions are limited within nanoscale. Such PNFs own some unique physical and chemical properties not only due to their highly porous structures but also because of the nanosize effects of their building blocks [1,2] and thus have potential applications in catalysts, electrodes, sensors, fuel/solar cells and batteries. Recently, our group proposed a new oxidation-assisted dealloying (OAD) [1] method to fabricate metal oxide PNFs whose building blocks are chips-like metal oxide nanoplates. During OAD of amorphous alloy films, it is expected that the corrosion starts with dissolution of defected or disordered less noble atoms [1] on the film surface which seem very instable and energetic. Nevertheless, during conventional dealloying of crystalline alloy ribbons, the corrosion starts with dissolution of a less noble atom on a flat alloy surface of closely-packed orientation [3]. Further, it has been shown that in conventional dealloying the microstructure of starting alloy ribbons such as crystalline [3,4], amorphous [5,6], single phase [7,8] or multiphase [4,7-9] affects the dealloying process and the resulting morphology to a certain extent. Similarly, we predict here that the microstructure of starting alloy films in OAD would also affect the OAD behavior and the resulting morphology in a certain unknown way. Thus, study on the influence of microstructure is suggestive and crucial to both the mechanism understanding and the technical controlling of PNF fabrication.

With the above considerations, in this paper we particularly studied the OAD of $Cu_{0.7}Al_{0.3}$ thin films with different microstructures which were obtained by high vacuum annealing at different temperatures. It was observed that different microstructures such as eutectic mixture or solid solution, the grain size of Cu or Al component in eutectic-mixture $Cu_{0.7}Al_{0.3}$ films affected the underlying OAD mechanism and the resulting morphology greatly. More important, it provided a new idea to fabricate flexible CuO PNFs with controllable pore size, porosity, block size and shape by changing the microstructure of starting Cu-Al alloy films.

## 2. Experimental

$Cu_{0.7}Al_{0.3}$ (at.%) films of ~134 nm in thickness were co-sputtered on well-cleaned 321 stainless steel (SS for short; Austenite, γ-Fe) substrates at room temperature in a JGP500A magnetron sputtering system. Other deposition details can be found in ref. [1]. As-deposited $Cu_{0.7}Al_{0.3}$ films were then annealed in a high vacuum of ~$7\times10^{-4}$ Pa for 30 min at different annealing temperatures to obtain alloy films with different microstructures. For simplicity, the SS substrate was marked as #0 and the SS-substrated samples were marked as #1 (as-deposited), #2 (annealed at 400 °C), #3 (annealed at 600 °C) and #4 (annealed at 800 °C) respectively.

In OAD, each of the as-deposited and annealed samples was immersed in a 10 mmol/L NaOH aqueous solution at room temperature under free corrosion conditions with a typical duration of 24 h. The OAD samples were repeatedly rinsed with deionized water and then kept in a vacuum chamber to avoid additional oxidation.

Morphological characterizations were made using field-emission scanning electron microscopy (FESEM, ZEISS SUPRA55). The detailed microstructures and compositions were characterized using powder X-ray diffraction (XRD, RIGAKU D/Max 2500 PC) and energy dispersive X-ray analysis (EDX) respectively.

## 3. Results and Discussion

Fig. 1(a) shows the XRD patterns of the SS-substrated as-deposited and annealed $Cu_{0.7}Al_{0.3}$ films and the corresponding SS substrate. It can be observed that the SS substrate (#0, in Fig. 1(a)) has a typical face-centered cubic (fcc) structure (JCPDS card no. 47-1417). The as-deposited $Cu_{0.7}Al_{0.3}$ film on SS substrate (#1, in Fig. 1(a)) shows a similar structure to the SS substrate only with a somewhat increase in diffraction intensity. It is clear that there are no diffraction peaks of Cu-Al and Al. Since the angular locations of diffraction peaks of Cu (JCPDS card no. 04-0836) are well consistent with that of SS substrate (JCPDS card no. 47-1417), the somewhat increase in diffraction intensity seems to indicate the existence of small Cu grains. Hence, the as-deposited $Cu_{0.7}Al_{0.3}$ film exhibits an amorphous Cu-Al eutectic mixture structure with some possible small Cu grains. When annealed at 400 ºC for 30 min, the $Cu_{0.7}Al_{0.3}$ film (#2, in Fig. 1(a)) displays a large increase in diffraction intensity at the angular locations of diffraction peaks of Cu. It indicates that Cu component tends to crystallize preferentially under high vacuum annealing. However, when annealed at a higher temperature of 600 ºC for 30 min, the $Cu_{0.7}Al_{0.3}$ film (#3, in Fig. 1(a)) presents a less increase in diffraction intensity with amplitude between #1 and #2 at the angular locations of diffraction peaks of Cu. Also, the diffraction peaks of Al are observed in the XRD pattern of #3. All the above demonstrates that as-deposited and annealed ($\leq 600\,^{\circ}$C) $Cu_{0.7}Al_{0.3}$ films uniformly show a eutectic mixture structure with separate crystallization of Cu and Al components. When annealed at 800 $^{\circ}$C for 30 min, however, a new solid solution structure of $Al_2Cu_3$ (JCPDS card no. 26-0015) is found in the $Cu_{0.7}Al_{0.3}$ film (#4, in Fig. 1(a)). Also, we can observe that the diffraction intensity at the angular locations of diffraction peaks of Cu is higher than that of #1 probably due to the existence of Cu grains.

Alternatively, one could also envision that high vacuum annealing may contribute to the increase of diffraction intensity of SS substrate. In fact, the heat treatment temperatures of 321 Austenite SS substrate and stabilizing treatment temperatures are often higher than 850 $^{\circ}$C. It is expected that the annealing below 800 $^{\circ}$C in this work would affect the SS substrate little. As a result, we can roughly get the relative intensity of Cu grains in as-deposited and annealed $Cu_{0.7}Al_{0.3}$ films by deducting the background SS substrate. As shown in Fig. 1(b), the Cu component after annealing especially in #2 is greatly crystallized. Relative to #2, however, the less increase of diffraction intensity of Cu at higher annealing temperature in #3 and #4 is probably due to the inter-inhibiting of two different crystallization processes, e.g. the crystallization of Cu and Al in #3, Cu and $Al_2Cu_3$ in #4.

In the following, the detailed corrosion morphologies of as-deposited and annealed $Cu_{0.7}Al_{0.3}$ films with different microstructures after OAD for 24 h in 10 mmol/L NaOH solution are shown in the FESEM images of Fig. 2.

Fig. 2(a) shows the corrosion morphology of as-deposited $Cu_{0.7}Al_{0.3}$ film (#1). A large-area and uniform porous nanoplate-film with an average pore size of 100 nm can be observed from the image. The building nanoplates with an average thickness of 20 nm and an average diameter of 300 nm are randomly inclined or perpendicular to the supporting substrate ("off-film" nanoplates). The surface of the nanoplates seems relatively compact and smooth. The FESEM image in Fig. 2(b) shows the corrosion morphology of $Cu_{0.7}Al_{0.3}$ film annealed at 400 °C (#2). However, different from Fig. 2(a), it displays a loose and porous nanoplate-film structure with an average pore size of 60 nm. The loose nanoplates with rough surfaces are comprised of some distinct nanoparticles around 20 nm in diameter. Fig. 2(c) shows the corrosion morphology of $Cu_{0.7}Al_{0.3}$ film annealed at 600 °C (#3). A nanoporous film can be found in the image, exhibiting a typical particle-void structure with average sizes of particles and voids both around 30 nm. Nevertheless, the corrosion morphology of $Cu_{0.7}Al_{0.3}$ film annealed at 800 °C (#4) shows not an obvious porous structure but a dense and flat film with some nanoplates on the film surface (see Fig. 2(d)). It demonstrates that $Cu_{0.7}Al_{0.3}$ films with eutectic mixture structures are suitable for the formation of CuO PNFs while $Cu_{0.7}Al_{0.3}$ films with solid solution structures seem not. Meanwhile, it also provides a new idea to fabricate flexible CuO PNFs with controllable pore size, porosity, block size and shape by changing the microstructure of starting Cu-Al alloy films.

By further checking FESEM images at lower magnifications, we also found some particles of micron-order in the corrosion products of #2 and #3. For example, Fig. 3 shows the typical corrosion morphology of #3 at a lower magnification relative to that of Fig. 2(c). Not only a PNF but also some micron-order particles on the film surface can be observed from the low-magnified image. To determine their chemical compositions, we performed EDX analysis at locations L1 (for PNF) and L2 (for micron-order particle) in Fig. 3 respectively. Their detailed composition reports are listed in Table 1, where elements C, Cr, Mn, Fe, and Ni come from the SS substrate while Cu, Al and O are mainly from the corrosion products. At location L1, the atomic ratio of Cu/O is about 1: 1.2, indicating that the composition of the PNF is CuO. The somewhat excessive O content (0.2) relative to Cu is probably from $Cr_2O_3$ on the surface of SS substrate or $Al_2O_3$ resulting from the oxidation of a little undissolved Al component (see Table 1). At location L2, the main component is Cu in spite of a little O content, indicating that the micron-order particles are almost of Cu grains. Similarly, further EDX analysis (not shown) demonstrates that the corrosion products in Fig. 2(a,b,d) are CuO except for some micron-order Cu particles in Fig. 2(b). We thus conclude that during OAD amorphous Cu component and small Cu grains can be easily oxidized into CuO while Cu grains of micron-order cannot.

As we all know, there is a large standard potential difference between Al and Cu (-1.662 V standard hydrogen electrode for $Al/Al^{3+}$ and 0.337 V for $Cu/Cu^{2+}$). So the less noble Al atoms in as-deposited and annealed $Cu_{0.7}Al_{0.3}$ films can be almost fully dissolved in NaOH solution while the undissolved more noble Cu atoms would further be oxidized, diffuse and self-assembly into CuO films [1]. Further, as will be presented in the following, different microstructures of starting $Cu_{0.7}Al_{0.3}$ films such as eutectic mixture/solid solution, amorphous/crystalline, or large/small grains would lead to different OAD processes and different morphologies of as-formed CuO films.

For as-deposited $Cu_{0.7}Al_{0.3}$ films (#1), it displays a eutectic mixture structure consisting of amorphous Cu-Al component along with some possible small Cu grains. The typical OAD process of #1 has been described detailedly in ref. [1] and can be summarized into four steps: (i) quick and collective dissolution of amorphous Al component and formation of a layer of Cu atoms without lateral coordination ("adatoms"); (ii) Cu "adatoms" are oxidized into $Cu_2O$ molecules by dissolved oxygen in electrolyte; (iii) $Cu_2O$ molecules diffuse about and agglomerate into $Cu_2O$ clusters; (iv) $Cu_2O$ clusters are further oxidized and self-assembly into CuO porous nanoplate-films due to the primary cell-induced oxygen consuming corrosion (see Eq. (1)) [1]. Since both of the starting Cu "adatoms" and Cu grains in #1 are small with sufficient mobility, as-formed CuO nanoplates display a compact and smooth surface (see Fig. 2(a)).

$$Cu^+ - e = Cu^{2+}, \varphi(Cu^+/Cu^{2+}) = 0.153\text{V}$$
$$\tfrac{1}{2}O_2 + H_2O + 2e = 2OH^-, \varphi(O_2/OH^-) = 0.348\text{V}$$
(1).

For $Cu_{0.7}Al_{0.3}$ films annealed at 400 ºC (#2), it displays another eutectic mixture structure which mainly consists of amorphous Al component and greatly crystallized Cu component. Different from #1, the Cu component in #2 exists in the form of Cu grains, each of which acts as a whole in the OAD process. Since large Cu grains are often with low mobility, the intermediate $Cu_2O$ clusters formed at the above step (iii) are too large to diffuse freely. As a consequence, as-formed CuO nanoplates present a rough and loose surface and look like some two-dimensional particle-aggregations (see Fig. 2(b)).

For $Cu_{0.7}Al_{0.3}$ films annealed at 600 ºC (#3), both Cu and Al components are crystallized. Similar to the case of #2, large Cu grains, each of which act as a whole with low mobility, would tend to form CuO particles or plate-like particle-aggregations. However, different from the amorphous Al component in #1 or #2, the dissolution velocity of crystalline Al component (Al grains) in #3 would be much slower. Thus, we can image the corrosion process of #3. Firstly, since the dissolution of Al component is nearly simultaneous from the film surface, similar to the case of #2, a layer of Cu grains would be left behind on the film surface after initial dealloying [1]. Secondly, as-formed Cu grains would be oxidized into $Cu_2O$ by dissolved oxygen and diffuse about along the film plane to form $Cu_2O$ clusters, rather than traveling into the film inside. This is because the dissolution velocity of Al grains is so low that maybe the next layer of the film has not been attracted or dissolved completely. Thirdly, different from the case of #1 or #2, there is still no sufficient $Cu_2O$ materials in the direction normal to the film surface which are required for the self-assembly of "off-film" CuO nanoplates. Thus, as-formed $Cu_2O$ on the film surface is further oxidized and preferentially agglomerates into CuO particles by primary cell-induced oxygen consuming corrosion mechanism [1]. In the following, the next layer has been totally dissolved and starts to follow the above OAD steps. In this way, the Cu component is oxidized and self-assembly into CuO particles layer by layer and thus the final corrosion products exhibit a typical particle-void structure (see Fig. 2(c)).

In #2 and #3, most of the Cu component exists in the form of Cu grains, each of which acts as a whole in the OAD process. Different size of Cu grains would not only cause the above mentioned different mobility but also lead to different oxidation resistance. For small Cu grains with large surface-to-volume ratio and high surface energy, as discussed above, they can be easily oxidized into $Cu_2O$ by the dissolved oxygen. Nevertheless, for Cu grains of micron-order, they can only be oxidized on the grain surface for several nanometers under free oxidation conditions [1]. During primary cell-induced oxygen consuming corrosion, the inner Cu core can be further oxidized in the following way:

$$Cu - 2e = Cu^{2+}, \varphi(Cu/Cu^{2+}) = 0.337V$$
$$\tfrac{1}{2}O_2 + H_2O + 2e = 2OH^-, \varphi(O_2/OH^-) = 0.348V$$
(2).

However, in this OAD process, the difference of electrode potentials between $Cu/Cu^{2+}$ and $O_2/OH^-$ is so small (0.011 V versus 0.195 V in Eq. (1)) that the oxidation of Cu grains would be very slow correspondingly. Thus, as demonstrated in Fig. 3 and Table 1, after 24-h OAD the micron-order grains in corrosion products is still of Cu component with little O. Note that in Eqs. (1-2) we use practical potentials of $Cu^+/Cu^{2+}$, $Cu/Cu^{2+}$ and $O_2/OH^-$ obtained by Nernst equation rather than standard potentials due to the non-standard conditions such as 10 mmol/L NaOH solution and 0.28 mmol/L dissolved oxygen (whose saturation value is about 8 mg/L under the conditions of 1 atm and room temperature [10]).

While for $Cu_{0.7}Al_{0.3}$ films annealed at 800 °C (#4), it displays a combination of a eutectic mixture structure (amorphous Al component and Cu grains) and a solid solution structure ($Al_2Cu_3$ grains). For the former, the OAD process is similar to that of #1 or #2 and thus CuO nanoplates are observed in the final corrosion product (see Fig. 2 (d)). While for the latter, similar to the case of #3, the dissolution of crystalline Al component in solid-solution $Al_2Cu_3$ grains is also very slow so that the more noble Cu "adatoms" would be oxidized and self-assembly into CuO particles layer by layer. Nevertheless, different from #2 and #3 where Cu grains act as a whole, single Cu atoms without lateral coordination ("adatoms") are left behind one by one and start to be oxidized, diffuse and agglomerate. Such Cu "adatoms" with high mobility can diffuse about freely and thus result in a compact and flat CuO film ultimately (see Fig. 2(d)). We should also note that, since the $Cu_{0.7}Al_{0.3}$ film is composed of many nanosized $Al_2Cu_3$ grains, the dissolution of Al component starting from the film surface would be frequently interrupted by the boundaries of grains whose closely-packed orientations are often not in the same direction (see Fig. 1, samples without ideal preferred orientation). In comparison, the starting alloy ribbons used in conventional dealloying are also of solid solution structures but typically with bulk grains. Such bulk grains can lead to a continuous dissolution of less noble atoms along the closely-packed orientation [3] and thus result in formation of bicontinuous ligament-channel structures.

It has been reported in literature that there are several practical applications of porous CuO or CuO-based composites. For example, Cao et al [11] and Jiang et al [12] developed porous $CuO-Fe_2O_3$ and $CuO/CeO_2$ composite catalysts respectively for carbon monoxide oxidation; Wang et al [13] synthesized porous CuO nanorods and Wan et al [14] prepared pillow-shaped porous CuO as anode materials for lithium-ion batteries; Samarasekara et al [15] found that porous CuO is a prime candidate in the application of carbon dioxide gas sensors and Hoa et al [16] synthesized porous CuO nanowires for hydrogen detection. It is expected that our prepared CuO PNFs with porous structures and nanoscale building blocks may also have potential applications in catalysts, electrodes and sensors. Specially, preliminary optical testing results in Ref. [1] have demonstrated a large blue shift of band gap from 1.2 eV to 2.01 eV for the porous CuO nanoplate-film in contrast to bulk CuO. Due to high absorption efficiency of ultraviolet ray and good transmittance of visible light (617 - 780 nm), such porous CuO nanoplate-film may be an ideal electrode candidate of solar cells.

## 4. Conclusions

In this paper, we studied the OAD of $Cu_{0.7}Al_{0.3}$ thin films with different microstructures which were obtained by high vacuum annealing at different temperatures. It was observed that different microstructures of starting $Cu_{0.7}Al_{0.3}$ films affected the corrosion morphology greatly. The as-deposited $Cu_{0.7}Al_{0.3}$ films at room temperature which showed an amorphous Cu-Al eutectic mixture structure were suitable for the preparation of CuO porous nanoplate-films. When annealed at 400 or 600 °C, the Cu or even Al component in $Cu_{0.7}Al_{0.3}$ films tended to crystallize separately which led to formation of CuO PNFs with building blocks of nanoparticles or plate-like nanoparticle-aggregations. However, when annealed at 800 °C, a solid solution structure of $Al_2Cu_3$ along with a eutectic mixture structure of Cu-Al was found in the $Cu_{0.7}Al_{0.3}$ films which seemed to form a dense and flat CuO film with some CuO nanoplates on the film surface. Also, the underlying mechanisms for the different corrosion morphologies of $Cu_{0.7}Al_{0.3}$ films with different microstructures were further explored. That is, large Cu grains with low mobility tended to form CuO particles or plate-like particle-aggregations, while crystalline Al component such as Al or $Al_2Cu_3$ grains with slow dissolution velocity would cause a layer-by-layer OAD. More important, it provided a new idea to fabricate flexible CuO PNFs with controllable pore size, porosity, block size and shape by changing the microstructure of starting Cu-Al alloy films. Thus, this study is suggestive and crucial not only to the mechanism understanding but also to the technical controlling of PNF fabrication.

**Figures**

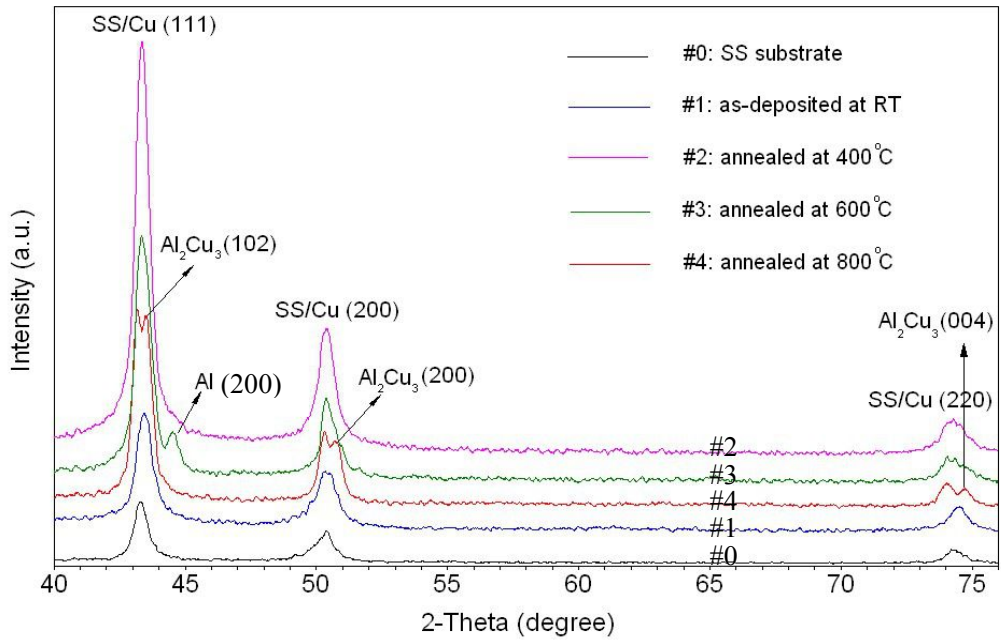

(a)

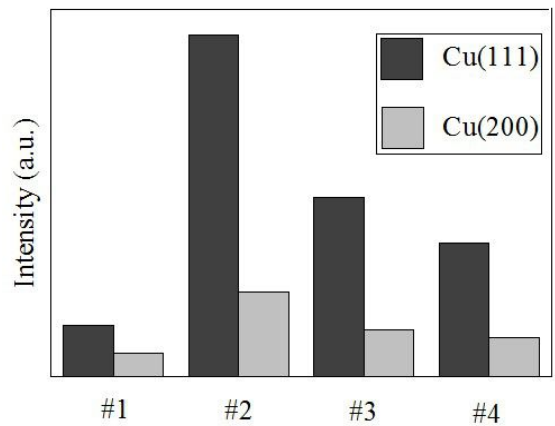

(b)

Fig. 1 (a) XRD patterns showing the microstructures of as-deposited and annealed $Cu_{0.7}Al_{0.3}$ films and the corresponding SS substrate; (b) The relative diffraction intensities of Cu grains in as-deposited and annealed $Cu_{0.7}Al_{0.3}$ films.

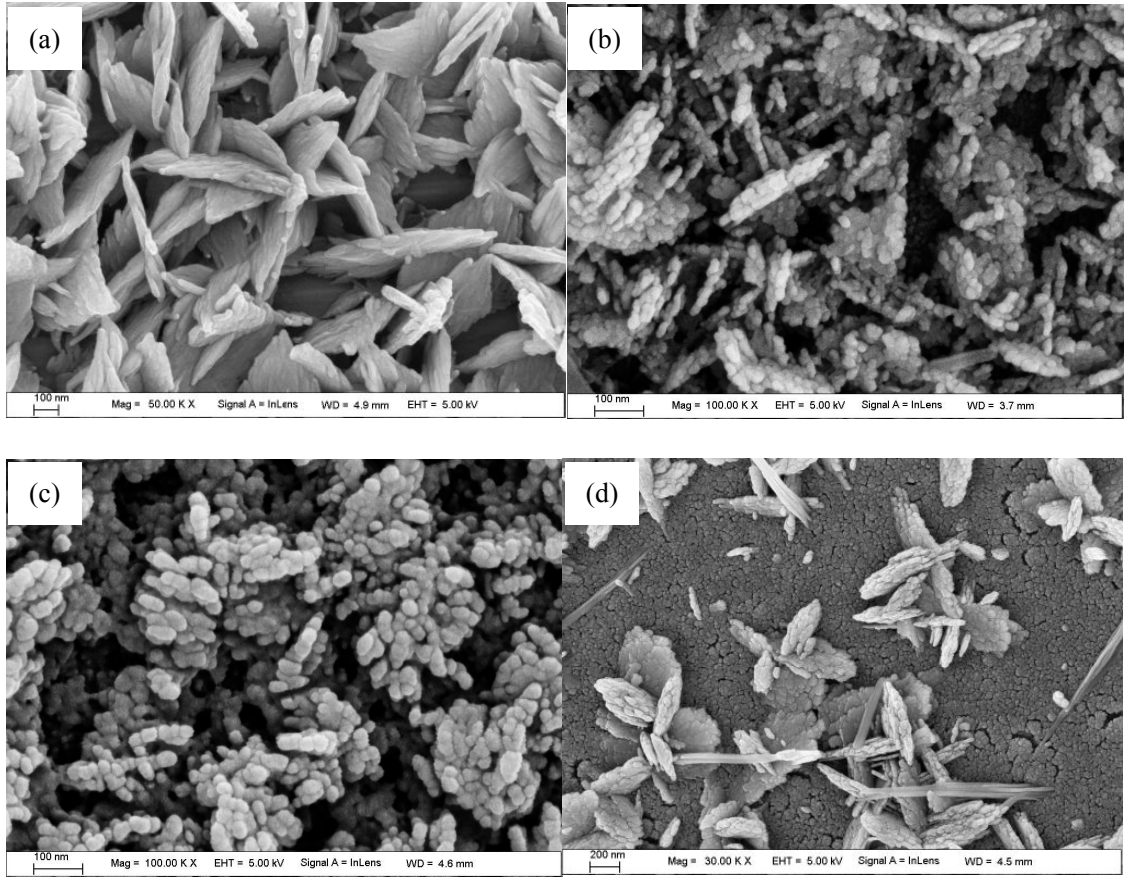

Fig. 2 High magnification FESEM images showing the detailed corrosion morphologies of as-deposited (a: #1) and annealed (b: #2; c: #3; d: #4) $Cu_{0.7}Al_{0.3}$ films after 24 h OAD in 10 mmol/L NaOH solution.

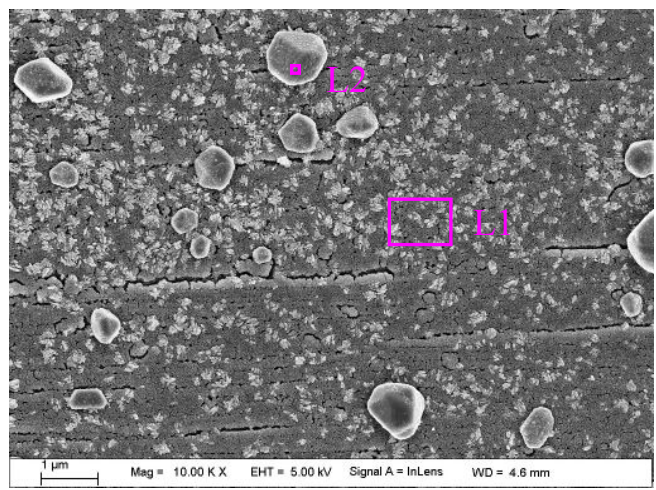

Fig. 3 FESEM image showing the corrosion morphology of #3 at a low magnification of 10,000×.

**Tables**

Table 1 Atomic percentage of elements at locations L1 and L2 in Fig. 3.

| Elements (at.%) | C | O | Al | Cr | Mn | Fe | Ni | Cu | Total |
|---|---|---|---|---|---|---|---|---|---|
| L1 | 39.59 | 17.28 | 1.97 | 5.52 | 0.27 | 18.98 | 1.93 | 14.46 | 100 |
| L2 | 50.24 | 2.42 | - | 0.45 | - | 1.71 | - | 45.19 | 100 |